\documentstyle[pre,aps]{revtex}

\begin{document}

\draft

\title{Cannon for Neutral Particles}

\author{V. I. Yukalov and E. P. Yukalova} 

\address{Bogolubov Laboratory of Theoretical Physics\\
Joint Institute for Nuclear Research, Dubna 141980, Russia}

\maketitle

\vspace{2cm}

\begin{abstract}

Dynamics of spin--polarized neutral particles, such as neutrons or neutral 
atoms and molecules, in magnetic fields is studied. A new regime of
motion is found where particles move mainly in one direction forming a 
well--collimated beam. This regime suggests a mechanism for creating
devices emitting directed beams of neutral particles.

\end{abstract}

\vspace{4cm}


\vspace{2cm}


The procedure of creating and accelerating directed beams of particles is 
of high practical importance. It would be difficult to enumerate various
applications of such directed beams. Charged particles, as is known, are
well manipulated by electric and magnetic fields [1]. To the contrary, for
neutral particles there are no known ways of creating and accelerating
directed beams by means of nonresonant electromagnetic fields. Directed
neutral beams have  been formed by employing mechanical collimators
selecting particles from an isotropic distribution, by accelerating neutral 
particles through long tubes with a high pressure difference between the
ends, as is done in molecular beam masers [2-4], and for resonance atoms
by using laser beams. In this paper we show that there exists a {\it
magnetic} mechanism that allows the formation of {\it directed} beams of
neutral particles with strong acceleration. This mechanism requires no
mechanical collimators, but does require a particular configuration of
magnetic fields and a special polarization of particle spins at the
initial time. A brief announcement of the results presented below has 
been given in Ref. [5].

To demonstrate the principal possibility of this new dynamical mechanism,
we consider a rarefied gas of particles whose collisions at the first
approximation can be neglected. Let particles be placed in magnetic fields
whose space variation is sufficiently smooth, so that the semiclassical
description can be applied [6]. Then for a particle with mass $m$ and
magnetic moment $\mu_0$, one may write the evolution equations for
quantum--mechanical averages. For the average real--space variable 
$\stackrel{\rightarrow}{R}=\{R^x,R^y,R^z\}$ one has 
\begin{equation}
\frac{d^2R^\alpha}{dt^2}=\frac{\mu_0}{m}\stackrel{\rightarrow}{S}
\cdot \frac{\partial\stackrel{\rightarrow}{B}}{\partial R^\alpha} \; ,
\qquad (\alpha=x,y,z)
\end{equation}
with initial conditions
$\stackrel{\rightarrow}{R}(0)=\stackrel{\rightarrow}{R}_0$ and
$\dot{\stackrel{\rightarrow}{R}}(0)=\dot{\stackrel{\rightarrow}{R}}_0$, 
where the dot means, as usual, a time derivative. It is easy to check [6]
that Eq. (1) follows from the definition of $\stackrel{\rightarrow}{R}$ 
as an average of a position operator with a wave function satisfying the
time--dependent Schr\"odinger equation. The average spin
$\stackrel{\rightarrow }{S}=\{S^x,S^y,S^z\}$ satisfies
the equation 
\begin{equation}
\frac{d\stackrel{\rightarrow }{S}}{dt}=\frac{\mu _0}{\hbar}
\stackrel{\rightarrow}{S}\times \stackrel{\rightarrow}{B},
\end{equation}
with an initial condition $\stackrel{\rightarrow}{S}(0)=
\{S_0^x,S_0^y,S_0^z\}$. Let us take the total magnetic field 
$\stackrel{\rightarrow}{B} =\stackrel{\rightarrow}{B}_1 +
\stackrel{\rightarrow}{B}_2$
as a sum of two terms,
\begin{equation}
\stackrel{\rightarrow}{B}_1 =
B_1'\left ( R^x\stackrel{\rightarrow}{e}_x + 
            R^y\stackrel{\rightarrow}{e}_y +
    \lambda R^z\stackrel{\rightarrow}{e}_z \right ),\;
\qquad
\stackrel{\rightarrow}{B}_2 = 
B_2 \stackrel{\rightarrow}{h}(t) ,
\end{equation}
the first being a quadrupole field, parametrized by its gradient $B_1'$
and the anisotropy parameter $\lambda$, and the second, a transverse
field, with
\begin{equation}
\stackrel{\rightarrow}{h}(t) = h_x\stackrel{\rightarrow}{e}_x +
h_y\stackrel{\rightarrow}{e}_y, \qquad |\stackrel{\rightarrow}{h}|= 1 ,
\end{equation}
depending only on time but not on real--space variables. Such magnetic
fields are easy to form and are often used in different applications. For
example, the quadrupole fields are the basis of quadrupole magnetic traps
and the Ioffe--Pritchard traps [7,8]. A transverse {\it rotating} field,
with $h_x=\cos\omega t$ and $h_y=\sin\omega t$, has also been employed in
magnetic traps [9]. This rotating field will be used below as a concrete
example. However, the effect we consider exists not solely for such a
rotating field but for a wide class of transverse fields satisfying the
condition
\begin{equation}
\left | d\stackrel{\rightarrow}{h} / dt \right | \ll
\mu_0 B_2 / \hbar \; .
\end{equation}
To be absolutely concrete, we also take the anisotropy parameter $\lambda=-2$ 
so that $\stackrel{\rightarrow}{\nabla}\cdot\stackrel{\rightarrow}{B}_1=0$. 
For what follows, it is convenient to measure the components of the space 
vector $\stackrel{\rightarrow}{R}$ in units of the characteristic length 
$R_0\equiv B_2{\Large /}B_1'$ defining the radius of the field zero in 
the radial direction. To this end, we define the dimensionless vector
$\stackrel{\rightarrow}{r}\equiv\stackrel{\rightarrow}{R}/R_0=\{ x,y,z\}$. 
Introduce the characteristic frequencies 
\begin{equation}
\omega_1^2 \equiv \mu _0B_1' / mR_0,\qquad
\omega_2 \equiv \mu_0B_2 / \hbar ,
\end{equation}
the first of which, $\omega_1$, corresponds to the motion of particles in
real space and the second, $\omega_2$, to the spin motion. The physical
meaning of these frequencies becomes evident after we present Eqs. (1) 
and (2) in the form
\begin{equation}
\frac{d^2\stackrel{\rightarrow}{r}}{dt^2}=
\omega_1^2\left( S^x\stackrel{\rightarrow}{e}_x + 
S^y\stackrel{\rightarrow}{e}_y -
2S^z\stackrel{\rightarrow}{e}_z\right) , \qquad
\frac{d\stackrel{\rightarrow}{S}}{dt}=
\omega_2\hat{A}\stackrel{\rightarrow}{S},
\end{equation}
where the matrix $\hat{A}=[A_{\alpha \beta }]$, with $\alpha ,\beta =1,2,3$,
consists of the elements $A_{\alpha\alpha}=0,\; A_{12}=-A_{21}=-2z,\;
A_{13}=-A_{31}=-y-h_y,\; A_{23}=-A_{32}=x+h_x$. From (7) it is really
clear that $\omega _{1}$ and $\omega _{2}$ are the characteristic
frequencies of the space and spin motions, respectively. 

Notice that the system of equations (7), with a nonuniform magnetic 
field, is invariant under the change $\stackrel{\rightarrow}{S}\rightarrow
-\stackrel{\rightarrow}{S}$ and $\stackrel{\rightarrow}{r}\rightarrow
-\stackrel{\rightarrow}{r}$. This invariance can be called the 
Stern--Gerlach symmetry since in the particular case of a uniform magnetic
field one would recover the conditions of the Stern--Gerlach experiment.

To solve the system of nonlinear differential equations (7), let us
recall that (1) and (2) are derived in the semiclassical approximation
whose criterion of validity is the slow space variation of magnetic fields
[6]. In our notation, this criterion can be written as the inequality 
$\left|\omega_1{\Large /}\omega_2\right| \ll 1$. This condition shows that
the space variable $\stackrel{\rightarrow}{r}$ can be treated as slow,
compared to the fast spin variable $\stackrel{\rightarrow}{S}$. From
inequality (5) it follows that $\stackrel{\rightarrow}{h}$ is also slow as
compared to $\stackrel{\rightarrow}{S}$. For the rotating field, condition
(5) simply means that $\omega\ll\omega_2$. Therefore, the system of nonlinear 
equations (7) can be solved by employing the method of scale separation 
[10-13] which is a variant of the Krylov--Bogolubov averaging method 
[14,15]. A detailed description of this approach as applied to the 
nonadiabatic dynamics of atoms in nonuniform magnetic fields has been 
given in Refs. [16,17]. Following the method of scale separation, we, 
first, need to solve the equation for the fast variable, keeping there 
the slow variables as quasi--integrals  of the motion. Then, under fixed 
$\stackrel{\rightarrow}{r}$ and $\stackrel{\rightarrow}{h}$, the second 
equation from (7) can be solved exactly. The resulting solution is 
\begin{equation}
\stackrel{\rightarrow}{S}(t) = 
\sum_{i=1}^3 a_i\stackrel{\rightarrow}{S}_i(t) , \qquad
a_i = \stackrel{\rightarrow}{S}(0)\cdot\stackrel{\rightarrow}{b}_i(0),
\qquad \stackrel{\rightarrow}{S}_i(t) =
\stackrel{\rightarrow}{b}_i(t)\exp\left\{ \beta_i(t)\right\} , 
\end{equation}
$$
\stackrel{\rightarrow}{b}_i(t) =\frac{1}{\sqrt{C_i}}\left [\left (\alpha_i
A_{13} + A_{12}A_{23}\right )\stackrel{\rightarrow}{e}_x +\left (\alpha_i
A_{23} - A_{12}A_{13} \right )\stackrel{\rightarrow}{e}_y +\left (\alpha_i^2
+ A_{12}^2\right ) \stackrel{\rightarrow}{e}_z \right ] , 
$$
$$
C_i =\left ( |\alpha_i|^2 - A_{12}^2\right )^2 +\left ( |\alpha_i|^2 +
A_{12}^2\right )\left ( A_{13}^2 + A_{23}^2 \right ) , \qquad
\alpha^2 = A_{12}^2 + A_{13}^2 + A_{23}^2\; , 
$$
$$
\alpha_{1,2}=\pm i\alpha , \qquad \alpha_3 = 0, \qquad
\beta_i(t) = \omega_2\int_0^t \alpha_i(t)\; dt\; , \qquad \beta_3(t) = 0 . 
$$
The fast solution (8) is to be substituted into the equation for the slow
variable, averaging the right--hand side of it over an interval of time
much longer than the period of fast oscillations $2\pi/\omega_2$, which
gives  $\langle\stackrel{\rightarrow}{S}\rangle =
a_3\langle \stackrel{\rightarrow}{b}_3\rangle$, and for the case of the
rotating field
\begin{equation}
a_3= \frac{(1+x)S_0^x + yS_0^y - 2zS_0^z}{[(1+x)^2 + y^2 + 4z^2]^{1/2}}, 
\qquad
\stackrel{\rightarrow}{b}_3 =
\frac{(x+\cos\omega t)\stackrel{\rightarrow}{e}_x +
(y +\sin\omega t)\stackrel{\rightarrow}{e}_y - 
2z\stackrel{\rightarrow }{e}_z}{[1 +2(x\cos\omega t + 
y\sin\omega t) + x^2 + y^2 + 4z^2]^{1/2}}. 
\end{equation}
In this way, we obtain 
\begin{equation}
\langle\omega_1^2\left( S^x\stackrel{\rightarrow}{e}_x + 
S^y\stackrel{\rightarrow}{e}_y - 
2S^z\stackrel{\rightarrow}{e}_z\right )\rangle = 
\frac{\omega_1^2\left[ (1 + x)S_0^x +yS_0^y - 2zS_0^z\right ]
\left( x\stackrel{\rightarrow}{e}_x + y\stackrel{\rightarrow}{e}_y +
8z\stackrel{\rightarrow}{e}_z\right )}
{2\left [ (1 + 2x + x^2 + y^2 + 4z^2)(1 +x^2 +y^2 +4z^2)\right ]^{1/2}}.
\end{equation}
As a result, we come to the equation 
\begin{equation}
\frac{d^2\stackrel{\rightarrow}{r}}{dt^2} =
<\omega_1^2(S^x\stackrel{\rightarrow}{e}_x + 
S^y\stackrel{\rightarrow}{e}_y -2S^z\stackrel{\rightarrow}{e}_z)>
\end{equation}
describing the averaged motion of particles. If we take adiabatic initial 
conditions for the spin polarization such that $S_0^x\neq 0$ and 
$S_0^y=S_0^z=0$, then for $\stackrel{\rightarrow}{r}\ll 1$, the 
right--hand side of Eq. (10) reduces to the simple harmonic force 
$\frac{1}{2}\omega_1^2 S_0^x(x\stackrel{\rightarrow}{e}_x + 
y\stackrel{\rightarrow}{e}_y + 8z\stackrel{\rightarrow}{e}_z)$. Hence Eq. 
(11) describes the adiabatic motion of particles in a harmonic potential. 
For $S_0^x<0$, this is the standard harmonic oscillations of trapped 
particles, while for $S_0^x>0$, the latter are not confined and escape 
from the trap in all directions. Neither of these known cases is of 
interest for us. Our aim is to find a principally different regime of 
motion, when the particles are neither completely trapped nor uniformly 
coupled out but move preferably in one direction forming a collimated 
beam. Assume that at $t=0$ the particles are prepared in the state with 
an initial spin polarization 
\begin{equation}
S_{0}^{x}=S_{0}^{y}=0,\qquad S_{0}^{z}=-S\; ,
\end{equation}
which is referred to nonadiabatic initial conditions [18]. The ways of 
preparing such polarized initial states are discussed in detail in Ref. 
[18]. To take into account the finite size of a device, we introduce the 
device form--factor, $\varphi(\stackrel{\rightarrow}{r})$, and use the 
notation 
\begin{equation}
f(\stackrel{\rightarrow}{r})=
\frac{\varphi(\stackrel{\rightarrow}{r})}
{[( 1 +2x +x^2 +y^2 +4z^2)(1 +x^2 +y^2 +4z^2)]^{1/2}}.
\end{equation}
Then the motion of particles is described by the equation
\begin{equation}
\frac{d^2\stackrel{\rightarrow}{r}}{dt^2} =
S\omega_1^2fz\left( x\stackrel{\rightarrow}{e}_x +
y\stackrel{\rightarrow}{e}_y +
8z\stackrel{\rightarrow }{e}_z\right ) .
\end{equation}
From here it is seen that the motion along the $x$ and $y$ axes is similar
to each other. One may also notice that Eq. (14) is invariant under the
inversion $z\rightarrow -z$ and $S\rightarrow -S$. Thence, the trajectories 
for $S>0$ are mirror--symmetric, with respect to the $x-y$ plane, to the
trajectories for $S<0$. Therefore, it is sufficient to study only a case,
say, when $S>0$, which will be assumed in what follows. It is also tempting 
to simplify the equations by passing to dimensionless time measured in
units of $(\sqrt{S}\omega_1)^{-1}$. For this purpose, we define 
$\tau=\sqrt{S}\omega_1t$. Thus, from (14), we come to the system of 
equations
\begin{equation}
\frac{d^2x}{d\tau^2} = fxz, \qquad \frac{d^2y}{d\tau^2} = fyz, \qquad
\frac{d^2z}{d\tau^2} = 8fz^2 .
\end{equation}

An analytic solution of Eq. (15) is possible only for the initial stage of
the motion, when $|\stackrel{\rightarrow}{r}|\ll 1$. The corresponding
solutions are given by the Weierstrass and Lam\'e functions [19,20]. From
the properties of these functions, it follows that the axial motion of
particles is bounded from below by the minimal value
$z_{min}=(z_0^3-\zeta)^{1/3}$, with $\zeta\equiv(3/16)\dot{z}_0^2$, and
$z_0\equiv z(0)$ being an initial axial position. So, particles can 
escape only in the positive $z$--direction. The motion along the axial 
direction is much faster than along the radial direction. This means that 
beam collimation begins already at the initial stage of the process. To 
analyze accurately the whole motion, we have to resort to numerical 
calculations. Let us specify the form--factor
$\varphi(\stackrel{\rightarrow }{r})$ assuming a spherical device with 
$\varphi(\stackrel{\rightarrow}{r})=
\exp(-|\stackrel{\rightarrow}{r}|^2{\Large /} L^2)$ , in which $L$ is a
characteristic size of the device, e.g., the radius of a coil forming
magnetic fields. The maximal velocity in the axial direction is 
$w_{max}\cong \left( \dot{z}_0 +2\sqrt{\pi}L\right)^{1/2}$. Therefore, 
taking $L$ sufficiently large, it is feasible to get arbitrary strong
acceleration in the axial direction. At the same time, the maximal
velocity in the radial direction, in the case of $L\gg 1$, is much less
than $w_{max}$, which can be accepted as the definition of collimation.

The results of numerical calculations are presented in characteristic
figures corresponding to the initial position at the center of the device
and to initial velocities varying in the interval $[-1,1]$ in all 
directions. All figures for different $L\gg 1$ are qualitatively the same,
because of which we fix here $L=1000$. Fig. 1 shows the trajectories of
particles in the  $x-z$ plane at the beginning of motion. It is clearly
seen how the particles having initially negative velocities in the axial
direction, after reaching the minimal value $z_{min}$, turn to the
positive $z$ direction. In Fig. 2 the trajectories are shown for longer
times, demonstrating how a well--collimated narrow beam is formed, being
stretched in the axial direction {\it more than an order of magnitude
stronger} than in the radial one. In Fig. 3 the velocities
$v(\tau)\equiv\dot{x}(\tau)$ and
$w(\tau)\equiv\dot{z}(\tau)$ are pictured, illustrating the acceleration
process in the axial direction. When varying $L$, all figures remain
qualitatively the same. The sole thing that changes is the scale on the
$z$--axis. Increasing $L$ by an order squeezes the $z$ scale approximately
twice in the figures with trajectories and three times in the figures with
velocities. Thus, the maximal velocity in the axial direction in the
dimensionless units is $w_{max}=6$ for $L=10$; $w_{max} =19$ for $L=100$;
and $w_{max}=60$ for $L=1000$, in accordance with the law $w_{max}\cong
(2\sqrt{\pi}L)^{1/2}$.

We would like to recall that equations (15) are derived from  (7) which is
identical to the initial Eqs. (1) and (2). In the derivation of (15) from
(7) the sole assumption used is the existence of the small parameters
$|\omega_1/\omega_2|\ll 1$ and $|\omega/\omega_2|\ll 1$. This made it
possible to apply the scale separation approach [10-13,16,17] whose 
mathematical foundation is based on the Krylov--Bogolubov averaging 
method [14,15]. 

In order to understand how to choose the characteristic parameters $B_1'$
and $B_2$ of the magnetic field (3), it is necessary to return to dimensional
units. For simplicity, we set $S\sim 1$. Then, the maximal $z$ velocity is
$w_{max}\cong (2\sqrt{\pi}L\mu_0 B_2/m)^{1/2}$. To achieve an effective
acceleration for a given sort of particles with fixed $\mu_0$ and $m$, we
should take $L\gg 1$ and sufficiently large $B_2$. However, increasing $L$
and $B_2$ implies the increase of the size of a device $l=LB_2/B_1'$.
Hence, to achieve $L\gg 1$ for a given device requires that $B_2\ll
lB_1'$. At the same time, the existence of the small parameter
$|\omega_1/\omega_2|\ll 1$ yields $\hbar^2(B_1')^2/m\mu_0B_2^3\ll 1$.
Therefore, for a given sort of particles and a given device we must have
$(\hbar^2/m\mu_0)^{1/3}(B_1')^{2/3}\ll B_2\ll lB_1'$. This fixes the
required relation between the parameters of the magnetic fields.  The
frequency of the rotating field, according to Eq. (5), is to be much
smaller than $\omega_2$.

The effect described in this paper is rather general, and choosing the
corresponding parameters, one could realize such a directed acceleration
for any kind of neutral particles having spins. To show that the parameters 
to be chosen are quite realistic, let us make numerical estimates for
$^{87}Rb$ in a magnetic trap of Ref. [9]. Then, the gradient of the 
quadrupole field is $B_1'=120\; G/cm$ and the amplitude of the rotating 
field is $B_2=10\; G$, with the rotating frequency $\omega=5\times 
10^4s^{-1}$. This gives $\omega_1\sim 10^2s^{-1}$ and $\omega_2\sim
5\times 10^7s^{-1}$. Hence, the inequalities $\omega_1\ll\omega\ll\omega_2$ 
hold true. The required small parameters $\omega_1/\omega_2\sim 10^{-6}$
and $\omega/\omega_2\sim 10^{-3}$ are really very small, because of which
the scale separation approach provides very accurate solutions differing
from the exact ones by negligible corrections. The radius of the atom
cloud is $R_0\sim 0.1\; cm$. Taking $L=10$, the characteristic radius of
coils would be  $l\sim 1\; cm$. For $L=100$, this would be $l\sim 10\; cm$. 
The maximal velocity $w_{max}\approx 60\; cm/s$ for $L=10$ and $w_{max}
\approx 200\; cm/s$ for $L=100$. Starting from an isotropic distribution
of velocities, $\dot{x}_0\sim\dot{z}_0$, as a result of the preferable
acceleration along the $z$ axis, one can obtain the radially squeezed
velocity diagram with a rather large squeezing factor. As the phase
portraits in Figs. 2 and 3 show, the squeezing factor is about $20$ for
velocities and $50$ for the real phase variables. Let us stress that such
a high degree of squeezing is achieved under a rather unfavorable
assumption of a spherical device with a spherically symmetric form--factor
$\varphi(\stackrel{\rightarrow}{r})$. Taking a cigar--shaped device would
strongly enhance the degree of collimation. It is also possible to achieve
additional squeezing of the beam taking a quadrupole field in 
Eq. (3) with essentially different field gradients along the axial and
radial directions. Thus, one can reach quite high degree of collimation
with a squeezing factor of $100$ or $1000$ and more. The realization of
the effect we described is quite feasible by using existing magnetic
traps. What one needs to do is to prepare particles in the initial state
with polarization (12). This could be achieved in several ways. For example,
one could prepare particles in the desired state inside one trap and then
quickly load them into another trap with the considered magnetic field 
[18]. Or it might be possible to form the necessary initial polarization 
by means of a short pulse.

The presented calculations are based on single--particle trajectories.
Many--particle physics, related to collisions of particles, has not been 
treated. Since we would like to stress the practical applicability of the 
considered mechanism, it is useful to give at least a coarse estimate on 
what happens if collisions are included. And the most important is to 
define conditions when the collisions of particles do not essentially 
disturb the single--particle picture and permit one to realize the 
semiconfining regime of motion, when particles move predominantly in one
direction.

In order to include particle collisions, we need to add to the 
right--hand side of the evolution equation (14) an additional term 
describing an effective force caused by these collisions. It is customary
to treat the collisional force as a random variable. To this end, we can 
model this force by a random vector $\gamma\stackrel{\rightarrow}{\xi}$,
where $\gamma$ is a collision rate and $\stackrel{\rightarrow}{\xi} =
\{ \xi_x,\xi_y,\xi_z\}$ is a stochastic vector variable. Following again the
common way, we may interpret the set $\{\xi_\mu(t)\}$, with $\mu=x,y,z$, 
as a set of Gaussian random variables characterized by the stochastic 
averages
\begin{equation}
\ll \xi_\mu \gg = 0 \; , \qquad \ll \xi_\mu(t)\xi_\nu(t')\gg =
2D_\mu\delta_{\mu\nu}\delta(t-t') \; ,
\end{equation}
in which $D_\mu$ is a diffusion rate in the $\mu$ direction. Adding the 
random collisional force to the right--hand side of the evolution 
equation (14), we have, instead of Eq. (15), the system of equations
$$
\frac{d^2x}{dt^2} = S\omega_1^2 f z x + \gamma\xi_x \; ,
$$
\begin{equation}
\frac{d^2y}{dt^2} = S\omega_1^2 f z y + \gamma\xi_y \; ,
\end{equation}
$$
\frac{d^2z}{dt^2} = 8 S \omega_1^2 f z^2 + \gamma\xi_z \; .
$$

As is evident, if particle collisions are intensive, so that the motion 
is dominated by the random collision terms, then no organized motion of 
particles coherently moving in one direction is possible. The semiconfining
regime can survive only if influence of collisions is sufficiently weak, 
so that the random terms in Eqs. (17) can be treated as perturbation. In 
such a case, the solutions to Eqs. (17) can be presented as
$$
x = x_1 + x_2 \; , \qquad y = y_1 + y_2 \; , \qquad z = z_1 + z_2\; ,
$$
where $x_1,\; y_1$, and $z_1$ are the solutions of the unperturbed Eqs. (15) 
and $x_2,\; y_2$, and $z_2$ are the solutions to the equations
$$
\frac{d^2x_2}{dt^2} = S\omega_1^2 f ( z_1 x_2 + x_1 z_2 ) + \gamma\xi_x \; ,
$$
\begin{equation}
\frac{d^2y_2}{dt^2} = S\omega_1^2 f ( z_1 y_2 + y_1 z_2 ) + \gamma\xi_y \; ,
\end{equation}
$$
\frac{d^2z_2}{dt^2} = 16 S\omega_1^2 f z_1 z_2  + \gamma\xi_z \; . 
$$
The unperturbed functions $x_1,\; y_1$, and $z_1$ can be considered as 
slow compared to the random functions $x_2,\; y_2$, and $z_2$. Keeping in 
Eqs. (18) the unperturbed functions as quasi--invariants, one obtains the 
solutions
$$
x_2(t) = \int_0^t G_x(t-t')\left [ \gamma\xi_x(t') +
S\omega_1^2 f x_1 z_2(t')\right ] \; dt' \; , $$
$$
z_2(t) = \int_0^t G_z(t-t') \gamma\xi_z(t') \; dt' \; , 
$$
where the solution $y_2(t)$, having the form similar to $x_2(t)$, is not 
written down and
$$
G_x(t) = \frac{{\rm sinh}(\omega_x t)}{\omega_x} \; , \qquad
G_z(t) = \frac{{\rm sinh}(\omega_z t)}{\omega_z} \; , \qquad
\omega_x = \sqrt{S\omega_1^2 f z_1}\; , \qquad \omega_z = 4\omega_x \; .
$$
According to the stochastic averages (16), we have
$$
\ll x_2(t) \gg = \ll z_2(t) \gg = 0 \; , 
$$
$$
\ll x_2^2(t) \gg = \frac{\gamma^2 D_xt}{\omega_x^2} \left [
\frac{{\rm sinh}(2\omega_x t)}{2\omega_x t} - 1 \right ] +
$$
$$
+ \frac{\omega_x^4x_1^2\gamma^2 D_zt}{\omega_z^2(\omega_z^2 -\omega_x^2)^2
z_1^2}\left\{ {\rm cosh}(\omega_x t){\rm cosh}(\omega_z t) +
\frac{{\rm sinh}(\omega_z t)}{\omega_z t} \left [ {\rm cosh}(\omega_z t)
- {\rm cosh}(\omega_x t)\right ] \right. -
$$
\begin{equation}
-\left. \frac{\omega_z}{\omega_x}
{\rm sinh}(\omega_x t){\rm sinh}(\omega_z t) - 1 \right\} \; ,
\end{equation}
$$
\ll z_2^2 \gg = \frac{\gamma^2 D_z t}{\omega_z^2} \left [
\frac{{\rm sinh}(2\omega_z t)}{2\omega_z t} - 1 \right ] \; .
$$
As is seen from here, the collisions will not disturb much the ordered 
motion of particles provided that
$$
\frac{\gamma^2 D}{\omega_1^3} \ll 1 \; , \qquad 
D \equiv \sup\{ D_x,\; D_y,\; D_z\} \; .
$$
If we take for estimates the collision rate as $\gamma\sim \hbar\rho 
a_0/m$, where $\rho$ is the density of particles and $a_0$ is a 
scattering length, and the diffusion rate as $D\sim k_B T/\hbar$, where 
$T$ is temperature, then we get the inequality
\begin{equation}
\frac{\hbar\rho^2 a_0^2 k_B T}{m^2\omega_1^3} \ll 1 \; .
\end{equation}
The latter shows that the influence of random particle collisions, disturbing 
the organized semiconfined motion, can be negligible if density, temperature,
or the scattering length are small enough to satisfy condition (20). When 
inequality (20) does not hold, the organized directed motion of particles 
will be essentially spoiled by collisions. Then the motion becomes more 
complicated, at the same time becoming of no interest for our purpose. 
Our aim here has been to find the conditions when the directed 
semiconfined motion of particles is possible. Such a regime looks like 
feasible since one always can satisfy condition (20) by varying the 
parameters of the system.

Concluding, we have advanced a novel general mechanism for creating
{\it well--collimated} beams of neutral particles by means of {\it
magnetic fields}. Such particles could be neutrons or neutral atoms and
molecules with nonzero spin. In particular, such a mechanism can be
employed for creating narrow beams of molecules in molecular--beam masers
or directed beams of neutral atoms for other purposes. The mechanism does 
not depend on statistics and can be used for Bose as well as for Fermi 
particles. Varying the magnetic field parameters and the shape of a
device, one can regulate the beam characteristics in wide limits,
achieving the desired degree of collimation.

\vspace{3cm}

{\bf Figure Captions}

\vspace{1cm}

{\bf Fig. 1}. The trajectories of particles at the initial stage of the
acceleration process, for $0\leq\tau\leq 5$.

\vspace{5mm}

{\bf Fig. 2}. The trajectories of particles for $0\leq\tau\leq 20$.

\vspace{5mm}

{\bf Fig. 3}. The velocities of particles in the radial, $v(\tau)$, and
axial, $w(\tau)$, directions for $0\leq\tau\leq 100$.

\end{document}